\documentclass[compsoc, conference, a4paper, 10pt, times]{IEEEtran}
\IEEEoverridecommandlockouts
\usepackage{wrapfig}
\usepackage{textcomp}
\usepackage{xcolor}
\usepackage{soul}
\usepackage{braket}
\usepackage{algorithmic}
\usepackage[ruled,vlined, linesnumbered]{algorithm2e}
\usepackage{graphicx}
\usepackage{cite}
\usepackage{amsmath,amssymb,amsfonts}
\usepackage{algorithmic}
\usepackage{graphicx}
\usepackage{textcomp}
\usepackage{xcolor}
\def\BibTeX{{\rm B\kern-.05em{\sc i\kern-.025em b}\kern-.08em
    T\kern-.1667em\lower.7ex\hbox{E}\kern-.125emX}}
\begin{document}

\title{
Obfuscating Quantum Hybrid-Classical Algorithms for Security and Privacy\\}

\author{\IEEEauthorblockN{Suryansh Upadhyay}
\IEEEauthorblockA{\textit{The Pennsylvania State University} \\
PA, USA \\
sju5079@psu.edu}
\and

\IEEEauthorblockN{Swaroop Ghosh}
\IEEEauthorblockA{\textit{The Pennsylvania State University} \\
PA, USA \\
szg212@psu.edu}}

\maketitle

\begin{abstract}

As the quantum computing ecosystem grows in popularity and utility, it is important to identify and address the security and privacy vulnerabilities before they can be widely exploited. One major concern is the involvement of third-party tools and hardware. 
As more companies, including those that may be untrusted or unreliable, begin to offer quantum computers as a service, the users will be motivated to use them especially if they are cheaper and readily available compared to trusted ones. This is primarily since the computing time on quantum hardware is expensive and the access queue is typically long. However, usage of untrusted hardware could present the risk of intellectual property (IP) theft. For example, the hybrid quantum classical algorithms like Quantum Approximate Optimization Algorithm (QAOA), that is popular to solve wide range of optimization problems, encodes the graph properties e.g., number of nodes, edges and connectivity in the parameterized quantum circuit to solve a graph maxcut problem. QAOA employs a classical computer which optimizes the parameters of a parametric quantum circuit (which encodes graph structure) iteratively by executing the circuit on a quantum hardware and measuring the output. For mission critical applications like power grid optimization, the graph structure can reveal the power grid and their connectivity (an IP that should be protected). The graph properties can be readily retrieved by analyzing the QAOA circuit by the untrusted quantum hardware provider. To mitigate this risk, we propose an edge pruning obfuscation method for QAOA along with a split iteration methodology. The basic idea is to, (i) create two flavors of QAOA circuit each with few distinct edges eliminated from the problem graph for obfuscation, (ii) iterate the circuits alternately during optimization process to uphold the optimization quality, and (iii) send the circuits to two different untrusted hardware provider so that the adversary has access to partial graph protecting the IP. Extra layers in QAOA circuit are added to recover the optimization quality degradation due to the proposed obfuscation. 
We demonstrate that combining edge pruning obfuscation with split iteration on two different hardware secures the IP and increases the difficulty of reconstruction while limiting performance degradation to a maximum of $10\%$ ($\approx5\%$ on average) and maintaining low overhead costs (less than 0.5X for QAOA with single layer implementation).

\end{abstract}


\section{Introduction}

\begin{figure}
    \centering
    \includegraphics[width= 3.25in]{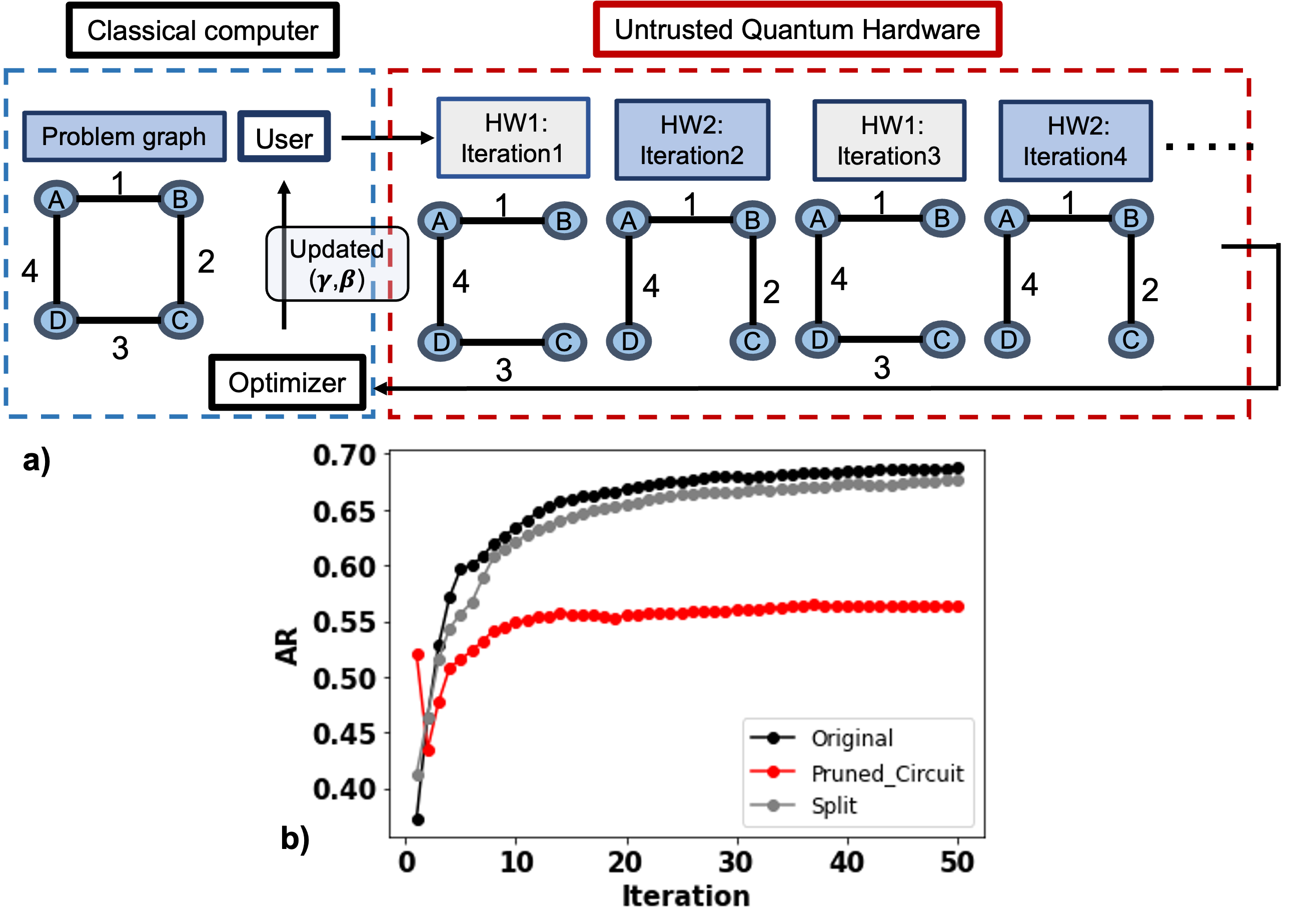}
    \caption{a) Two different versions of pruned graphs are obtained from the problem graph and run on two different untrusted hardware as partial circuits for alternative iteration. b) Sample 4-node maxcut using QAOA, simulated on the fake back-end (Fake\_montreal) for 50 iterations. The performance (Approximation Ratio) of the obfuscated circuit is significantly degraded when an edge is removed (pruned), however we report performance recovery using the proposed split iteration heuristic.}
    \label{1}
\vspace{-4mm}
\end{figure}

Quantum computing (QC) has the potential to solve many combinatorial problems exponentially faster than classical counterparts by utilizing the quantum-mechanical properties such as, superposition and entanglement. It can be used in various fields such as, machine learning \cite{b1}, security \cite{b2}, drug discovery\cite{b3}, and optimization \cite{b20}. However, QC also faces technical challenges like qubit decoherence, measurement error, gate errors, and temporal variation which can lead to errors in the output of a quantum circuit. While quantum error correction codes (QEC)\cite{b5} can provide reliable operations, they currently require a large number of physical qubits per logical qubit, making them impractical for widespread use in the near future. Noisy Intermediate-Scale Quantum (NISQ) computers, which have a limited number of qubits and operate in the presence of noise, offer a potential solution to important problems such as discrete optimization and quantum chemical simulations. To make the most of the limited capabilities of these computers, various hybrid algorithms have been proposed, such as the Quantum Approximate Optimization Algorithm (QAOA) \cite{b16} and Variational Quantum Eigensolver (VQE) \cite{b4}, in which a classical computer iteratively adjusts the parameters of a quantum circuit to guide it to the best solution for a given problem. 


\textbf{Motivation:} When using variational algorithms such as, QAOA to design problem-specific parametric quantum circuits to solve certain problems, the topology of the problem is embedded in the circuit and can be considered as an asset or intellectual property (IP) (refer to Fig. \ref{3}). For example, in applications such as, power grid or other critical infrastructure optimization, the client may want to keep the problem information confidential. These IPs may not present a risk for small scale quantum circuits that can be compiled on trusted vendors such as, IBM and Rigetti. However, with the emergence of third-party service providers offering potentially higher performance, and the scaling trend of current Noisy Intermediate Scale Quantum (NISQ) computers there is an increased risk of IP infringement. Third-party compilers like Orquestra \cite{b8} and tKet \cite{b9} that support hardware from multiple vendors are becoming available. Additionally, companies such as Baidu \cite{b38}, the Chinese internet giant, have recently announced all-platform quantum hardware-software integration solutions, such as Liang Xi, that provide access to various quantum chips via mobile app, PC, and cloud, and connect to other third-party quantum computers. These trends not only lead to reliance on untrustworthy third-party compilers and hardware suites instead of trusted counterparts, but also to reliance on third-party service providers, which can pose a significant risk to IP protection as these hardware providers/compilers can steal sensitive intellectual property (IP) and problem properties. It is crucial to ensure that IP is protected in the quantum computing ecosystem through robust security measures and legal protection.

To mitigate these risks, this paper proposes an obfuscation approach to protect IP while ensuring correct compilation and functionality. \emph{To the best of our knowledge, this is the first effort to identify a new security and privacy threat space for hybrid-classical quantum algorithm QAOA and develop countermeasures}.

\textbf{Proposed Idea:} QAOA can be used to solve complex optimization problems, such as the graph maxcut. The QAOA circuit is composed of two main parts: the mixing unitary and the problem unitary. The problem unitary is responsible for encoding the problem into the quantum circuit, while the mixing unitary is used to mix the different states of the circuit together. In our case, we consider the maxcut problem, used for finding the maxcut of a graph. To protect the IP of the QAOA circuit, one can remove or add information (e.g., graph edges) in the original graph. For example, considering a 4-node graph where nodes A, B, C, and D are connected in a circular fashion (Fig. \ref{1}a). 
For this case, if an edge (A, B) is removed the adversary will have 3 possible edges to check to determine the original graph, (A,C), (A,D), and (B,D), which would take $2^3 = 8$ trials in the worst-case scenario. Furthermore, there is a lack of oracle model (the quantum chip implements functionality using microwave/laser pulses that are not publicly accessible) to validate the adversarial guess. Therefore, adversarial effort to RE the obfuscated design is high. Any attempt to reuse the circuit without adding the removed edge will result in corrupted or severely degraded performance. As an example, Fig. \ref{1}b illustrates the performance degradation when only the pruned circuit is used. An incomplete circuit will not be able to generate all possible solutions, leading to suboptimal solutions. Furthermore, using an incomplete circuit for QAOA can result in poor performance as it may get stuck in a local optima. Our approach eliminates the information from the problem graph without degrading the solution quality. This is achieved by, (i) developing two variations of QAOA circuit by removing certain distinct edges from the problem graph to conceal the IP , (ii) iterating the circuits alternately during optimization process to uphold the optimization quality, (iii) sending the circuits to two different untrusted hardware provider so that the adversary has access to partial graph protecting the IP. The proposed heuristic ensures that each hardware device only receives a portion of the circuit, making it challenging for an attacker to reverse engineer the complete circuit. At the same time, the optimization performance is retained. To further illustrate the idea , we consider a simple example of a 4-node graph with only adjacent edges connected, specifically, with 4 edges (E0, E1, E2, E3). We apply the QAOA algorithm to three different cases: the original circuit, an obfuscated circuit with one edge (E0) removed, and an obfuscated circuit with the proposed hardware split heuristics, each for a total of 50 iterations Fig. \ref{1}. We use the Approximation Ratio (AR)(Section 2.6.2) as a metric to evaluate performance. The performance of the obfuscated circuit is significantly degraded when an edge (E0 out of 4) is removed, while leaving the adversary with limited information to determine the original circuit. However, when we run the first circuit (missing E0) on hardware 1 and the second circuit (missing E1) on hardware 2 with alternating iterations, we are able to recover performance while ensuring that each hardware device only has access to a partial circuit, making it challenging for an attacker to reverse engineer the full circuit. 

\textbf{Contributions:} We, (a) propose a novel threat model, (b) propose hardware split heuristic along with edge pruning obfuscation technique to counteract the threat, (c) perform exhaustive experiments on graphs of varying complexity, (d) present analysis and validation of the proposed heuristic.

\textbf{Paper organization:} In the remaining paper, Section II  provide quantum computing background and related work. The proposed threat model is described in Section III. Section IV describes the proposed obfuscation procedure and presents the simulations, results and analysis. The discussions are presented in Section V. Section VI concludes the paper.

\section{Background}

In this section, we discuss the basics of a quantum computer and the terminologies used in this paper.

\subsection{Qubits}

Qubits are analogous to classical bits that store data as various internal states (i.e., $\ket{0}$ and $\ket{1}$). In contrast to a classical bit, which can only be either 0 or 1, a qubit can concurrently be in both $\ket{0}$ and $\ket{1}$ due to quantum superposition. Hence, while a standard n-bit register can only represent one of the $2^n$ basis states, an n-qubit system can represent all $2^n$ basis states concurrently. A qubit state is represented as  $\varphi$ = a $\ket{0}$ + b $\ket{1}$ where a and b are complex probability amplitudes of states $\ket{0}$ and $\ket{1}$ respectively. Qubits are frequently visualized as a point on the so-called Bloch Sphere. On measurement, the qubit is reduced to a single state, i.e., a pure state of $\ket{0}$ or $\ket{1}$  with probability of $|a|^2$ and $|b|^2$ respectively. Qubit states can also be entangled by correlating the states of two or more qubits. The states of other entangled qubits can be changed by performing a single operation on one of the entangled qubits. 

\subsection{Quantum Gates}

In quantum systems, computation is performed by manipulating the qubit states. To produce the desired output, the gate operations change the amplitudes of the qubits. A quantum program performs a series of gate operations on a group of correctly initialized qubits (using laser pulses in ion trap qubits and RF pulses in superconducting qubits). Mathematically, quantum gates are represented using unitary matrices (a matrix U is unitary if U$U^\dagger$ = I, where U$^\dagger$ is the adjoint of matrix U and I is the identity matrix). For an n-qubit gate, the dimension of the unitary matrix is 2n×2n. Any unitary matrix can be a quantum gate. Only a few gates, known as the quantum hardware's native gates are currently practical in current systems. ID, RZ, SX, X (single qubit gates), and CNOT (2-qubit gate) are the basic gates for IBM systems. 

\begin{figure}
    \centering
    \includegraphics[width= 3.25in]{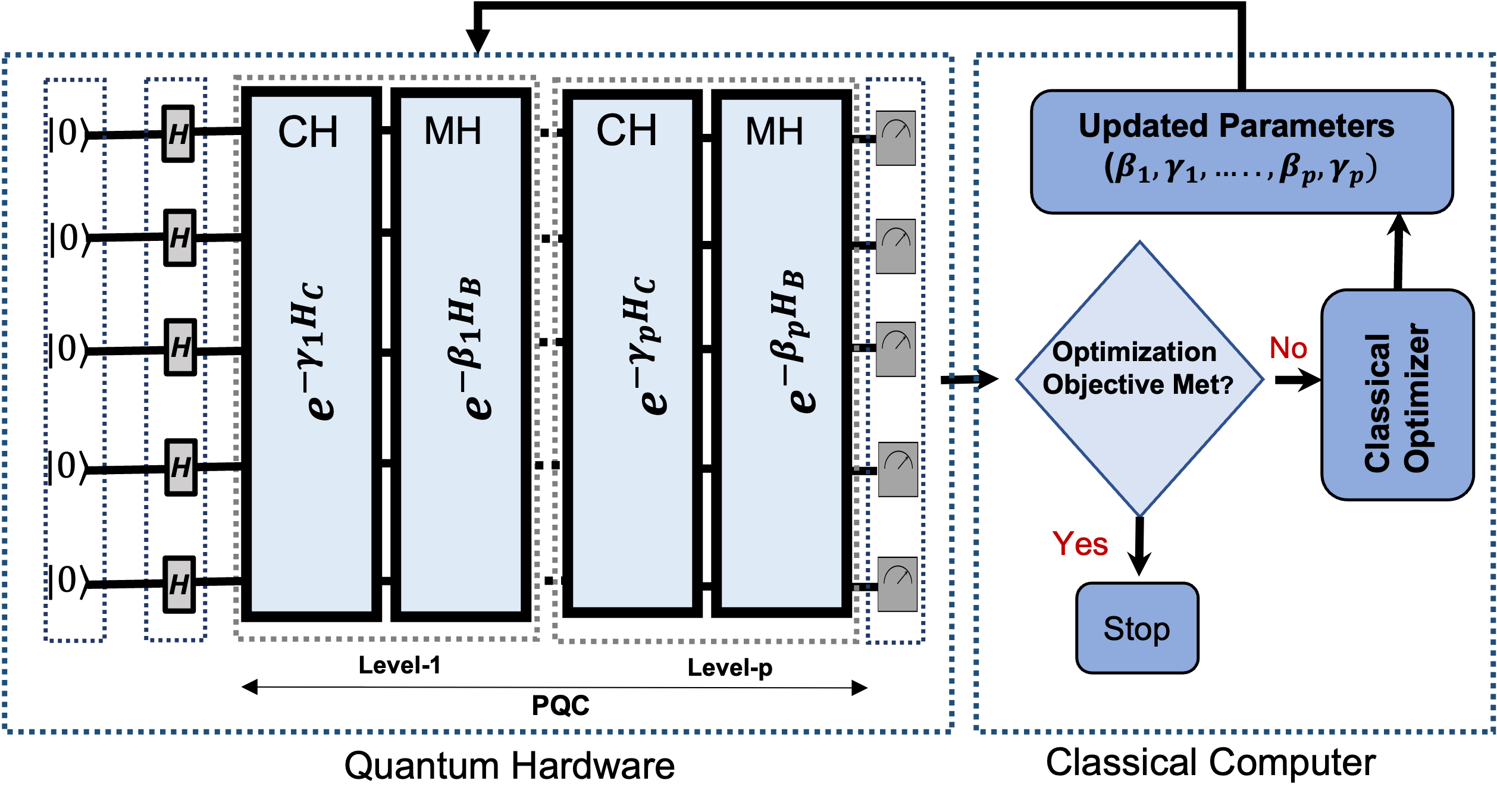}
    \caption{Schematic of a p-level quantum-classical hybrid algorithm QAOA. A quantum circuit takes input qubit states and alternately applies Cost Hamiltonian (CH) which encodes the problem and Mixing Hamiltonian (MH) `p' times and the final state is measured to obtain expectation value with respect to the objective function. This is fed to a (classical) optimizer to find the best parameters $(\gamma, \beta)$ that maximizes or minimizes the cost.
    }
    \label{2}
\vspace{-4mm}
\end{figure}

\subsection{Quantum program and instruction sets}

Similar to classical computing a quantum program can be represented in the well-adopted quantum circuit model\cite{b29} such that it may be characterized as a set of quantum gates operating on qubits to converge the output to a particular solution. A quantum program is composed of logical qubit variables and quantum operations that can modify the state of the qubits. Higher-level algorithms are converted into physical instructions that can be executed on quantum processors using quantum instruction sets. There are numerous instruction set architectures available, such as cQASM, OpenQASM, Quil, Blackbird, and others.

\subsection{Quantum error}

 Quantum gates are realized with pulses that can be erroneous. Due to variation, the pulse intended for a particular angle rotation may under-rotate or over-rotate, leading to erroneous logical operation. Quantum gates are also prone to error due to noise and decoherence \cite{b10}. Hence qubits interact with their surroundings and lose their states making the output of a quantum circuit error-prone. The deeper quantum circuit needs more time to execute and gets affected by decoherence which is usually characterized by the relaxation time (T1) and the dephasing time (T2). The buildup of gate error \cite{b11} is also accelerated by more gates in the circuit. Cross-talk is another type of quantum error in which parallel gate operations on multiple qubits can degrade each other's performance. Because of measurement circuitry flaws, reading a qubit with a 1 may result in a 0 and vice versa. The frequency of these errors differs between qubits and hardware, and they may be used as a mark to distinguish one piece of hardware from another.

\subsection{Cloud-based quantum backends}

Currently, IBM, Google, Microsoft, Qutech, QC Ware, and AWS Braket are among the top cloud vendors that provide users with access to quantum hardware (both superconducting and Trapped Ion qubits). However, access to quantum computers is prohibitively expensive, prompting users to seek out less expensive quantum hardware to solve their problems. In the future, quantum computers may be available from untrustworthy third parties (that may be located offshore). For example, third-party compilers, such as Orquestra \cite{b8} and tKet \cite{b9}, that are capable of supporting hardware from multiple vendors are becoming increasingly available. Furthermore, Baidu, a Chinese internet giant, have recently announced all-platform quantum hardware-software integration solutions, Liang Xi\cite{b38}. Untrusted hardware can steal sensitive IPs embedded in the quantum circuit. This is very similar to the situation with untrusted compilers which may produce very optimal circuits but may steal sensitive IP\cite{b32}. 
 
\subsection{Quantum algorithms}

Quantum algorithms are programs that are fundamentally quantum or exploit some important characteristic of quantum computation such as quantum superposition or entanglement solving some problems substantially quicker. They operate on a realistic model of quantum computation (the most frequent being the quantum circuit model \cite{b29}). Initially quantum computing algorithms were designed with fault tolerant quantum computer in mind, with the quantum gate model studied largely without noise \cite{b30}. Shor's algorithm for factoring and Grover's algorithm for searching an unstructured database or an unordered list are two of the most well-known algorithms. 

Commercially available now are Noisy Intermediate-Scale Quantum (NISQ) computers, which are sensitive, noisy, and error-prone. NISQ devices cannot effectively implement quantum algorithms such as Shor's algorithm because to their limited qubit capacity. However, a number of quantum-classical hybrid algorithms \cite{b21, b22, b23} based on variational principles have been developed to demonstrate quantum supremacy and make the best use of currently available near-term machines. By combining the capabilities of a quantum and a classical computer, these hybrid algorithms accelerate a task faster than its fully classical counterpart.

\begin{figure*}
    \centering
    \includegraphics[width= 6.75in]{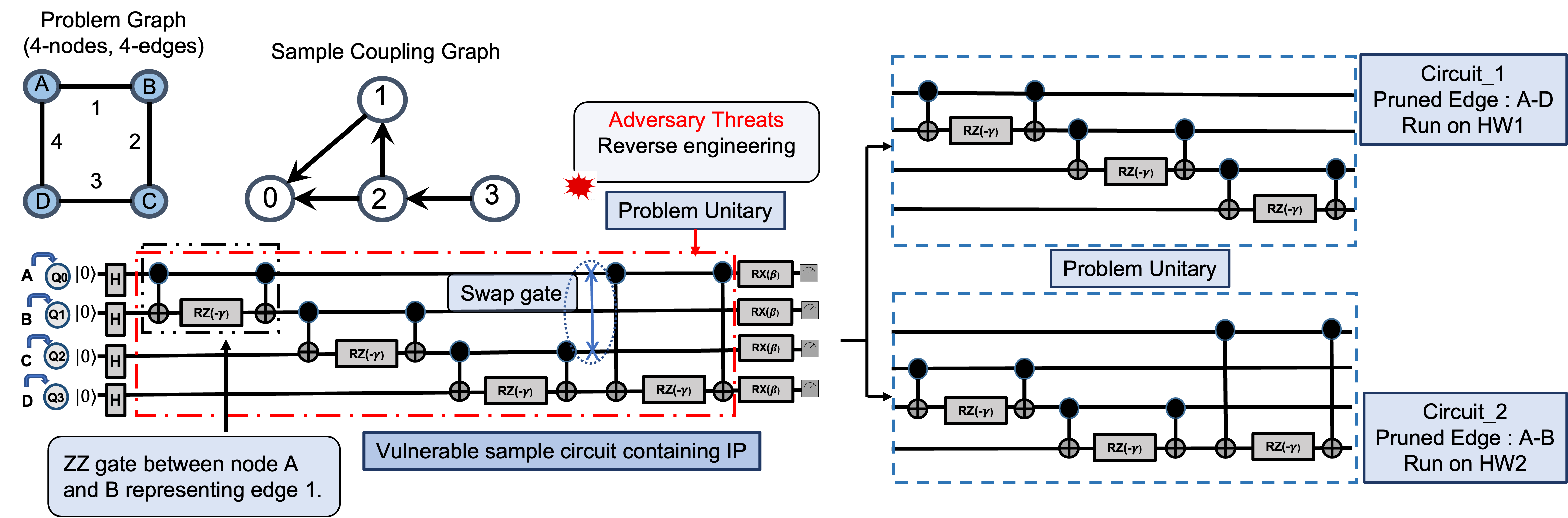}
    \caption{Threat model proposed in this paper. QAOA is used to solve max-cut for a sample 4 node graph, the problem-specific information is encoded in the problem unitary. The pruned circuits are sent to the different hardware providers eliminating one edge (edge specific gates from the problem unitary) for hardware 1 (Edge: A-D) and including that edge but removing another for hardware 2 (Edge: A-B),  ensuring each hardware device only gets a partial circuit, making it difficult for an attacker to reverse engineer the full circuit.}
    \label{3}
\vspace{-4mm}
\end{figure*}

\subsubsection{Parametric quantum circuit (PQC)}

PQC is built from a collection of parameterized and controlled single qubit gates. A classical optimizer optimizes the parameters iteratively to achieve the desired input-output relationship. A quantum processor employs a PQC to prepare a quantum state. The output distribution generated by repeatedly measuring the quantum state is then fed into a classical optimizer. The classical computer generates a new set of optimized parameters for the PQC based on the output distribution, which is then fed back to the quantum computer. The entire procedure continues in a closed loop until a traditional optimization target is reached. In recent years, quantum routines that are inherently resilient to errors have been developed using PQC \cite{b20, b33, b34}. 

\subsubsection{QAOA}

QAOA is a hybrid quantum-classical variational algorithm designed to solve combinatorial optimization problems. The quantum state in QAOA is created by a p-level variational circuit with 2p variational parameters. Even at the smallest circuit depth (p = 1), QAOA delivers non-trivial verifiable performance guarantees, and the performance is anticipated to get better as the p-value increases \cite{b14}, however the study \cite{b18} demonstrates that noise sources place a limit on those claims and the anticipated improvement. Fig.\ref{2} shows an overview of QAOA to solve a combinatorial problem. Recent developments in finding effective parameters for QAOA have been developed \cite{b14,b15,b16,b17}. 

In QAOA, a qubit is used to represent each of the binary variables in the target cost function C(z). In each of the p levels of the QAOA circuit, the classical objective function C(z) is transformed into a quantum problem Hamiltonian (Fig.\ref{2}). The output of the QAOA instance is sampled many times with optimal control parameter values, and the classical cost function is evaluated with each of these samples. The solution is determined by the sample measurement with the highest cost\cite{b19}. In a quantum classical optimization procedure, the expectation value of $H_C$ is determined in the variational quantum state $ E_p(\gamma,\beta)= \varphi_p(\gamma, \beta)|H_C|\varphi_p(\gamma, \beta)$. A classical optimizer iteratively updates these variables $(\gamma, \beta)$ so as to maximize $E_p(\gamma, \beta)$. A figure of merit (FOM) for benchmarking the performance of QAOA is the approximation ratio (AR) and is given as \cite{b14}
\begin{equation}\label{eq:AR}
    AR = E_p(\gamma, \beta)/Cmax
\end{equation}
where $Cmax = MaxSat(C(z))$.

\subsection{Relation to prior work}

In \cite{b32}, the authors proposed a method for protecting IP from an untrusted compiler by adding dummy gates in quantum circuits for obfuscation. This makes it difficult for an adversary to extract the original circuit without removing the dummy gates, which is a computationally hard problem. The authors developed a heuristic to find an optimal location for the dummy gate insertion that causes significant degradation in the output. Another work, \cite{b36} proposes split compilation to address the same issue. The idea is to split the quantum circuit into multiple parts that are sent to a single compiler at different times or to multiple compilers. In contrast, we propose a new approach for protecting IP embedded in hybrid-classical algorithms from untrusted quantum hardware. 
The works in \cite{b41}\cite{b42} focus on trojan insertion in reversible circuits before fabrication, which is not applicable to quantum circuits since they are not physically fabricated. Another work \cite{b43} assumes an untrusted foundry that can locate ancillary and garbage lines in a reversible circuit and extract the circuit functionality. Dummy ancillary and garbage lines are added to the circuit which increases the ancillary and garbage lines post-synthesis. However, this approach is only applicable for oracle-type or pure Boolean logic based quantum circuits and not for general quantum computing. 

Authors in \cite{b2} assume that a malicious adversary in the form of untrusted hardware provider will report incorrect measurement results to the user, thereby tampering with the results. They model and simulate adversarial tampering of input parameters and measurement outcomes on QAOA. Whereas our adversarial model considers the hardware provider to be untrusted with the objective to steal the IP embedded within the quantum circuit. Therefore, our proposed approach aims to provide a more robust solution for protecting IP in hybrid-classical algorithms such as, QAOA. 

\section{Threat Model}

In this section, we describe the attack model, adversary capabilities and feasibility of the proposed threat.

\subsection{Adversary capabilities}

We assume that adversary, (a) has access to the circuit run by the user. This is likely if the quantum computing cloud provider is rogue, (b) has the expertise on quantum computing principles e.g., construction of QAOA circuit for a given problem and knows the details of various qubit technologies, (c) can reverse engineer the problem properties such as, the graph from the QAOA problem PQC, (d) does not manipulate the quantum circuit, as the aim of the adversary is to steal the IP.

\subsection{Reverse engineering attack }

A quantum circuit may contain sensitive IPs such as the problem being solved, financial analysis, and proprietary algorithms that must be safeguarded. One potential threat to this IP is the use of untrusted third-party hardware providers. The adversary in the proposed scenario takes the form of a less reliable/untrusted quantum service provider who may pose as a reliable or trusted hardware provider. 
These attacks can be carried out by a variety of actors, including nation-states, cybercriminals, and competitors in various industries. The goal of a reverse engineering attack on QAOA is to gain information about the original optimization problem. For example: In the field of portfolio management, an attacker could use this information to learn about the portfolio of the user; In 
power grid optimization, the attacker could learn about the number and connectivity of power nodes which could be used to fine-tune or craft a follow up attack on the power grid. 

\textit{Reverse engineering the graph from the QAOA}: In QAOA circuit for solving graph maxcut problem, each node in the graph corresponds to a qubit and each edge corresponds to a ZZ gate applied between the qubits represented by the nodes (Fig. \ref{3}). Therefore, recovery of the graph from the QAOA circuit could be trivial. A potential complexity in the RE process could be due to swap gates that are added by the compiler to meet the coupling constraints of the target hardware. This is specifically true for superconducting qubit technology. For example, consider a coupling map as shown in Fig. \ref{3}, where the qubits are arranged in a two-dimensional grid and can interact with their nearest neighbors. To apply a ZZ gate between qubit $Q_0$ and qubit $Q_3$, a swap operation would be needed to first bring the two qubits together. This can be achieved by swapping the positions of $Q_2$ and $Q_3$. An adversary may infer the swap gates (since they contain sequence of 3 CNOT gates) in the QAOA circuit, remap the physical to logical qubit and still reconstruct the original graph. 

\begin{figure}
    \centering
    \includegraphics[width= 3.25in]{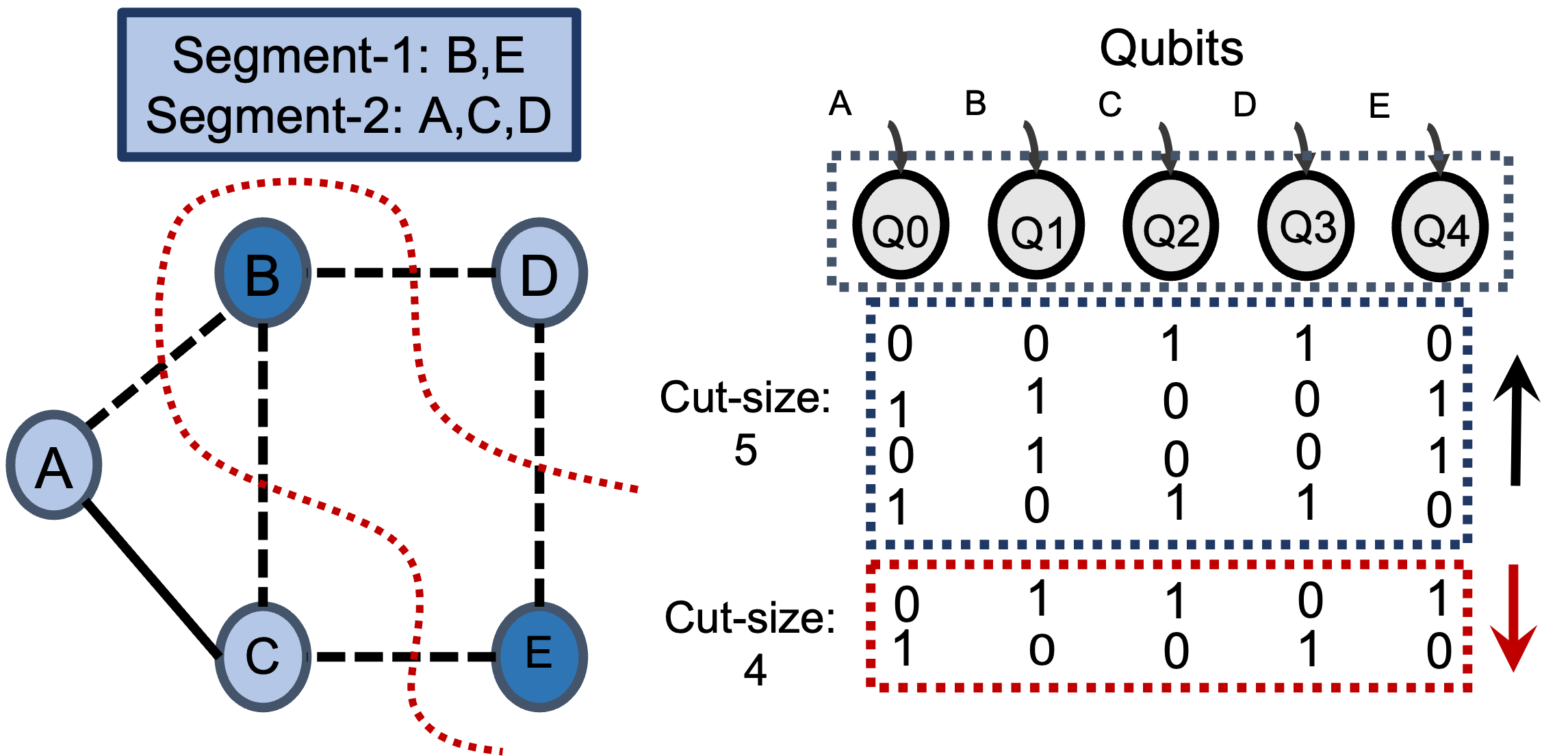}
    \caption{QAOA illustration for solving the maximum-cut (MaxCut) problem. The MaxCut size for the represented 5 node graph is 5 (cut in red). The probabilities of basis state measurements that represent larger cut-sizes for the problem graph are increased iteratively by QAOA. After the QAOA is completed, the states in cut size 5 will have higher probabilities than the states in cut size 4.
    }
    \label{4}
\vspace{-4mm}
\end{figure}

\begin{figure}
    \centering
    \includegraphics[width= 3.25in]{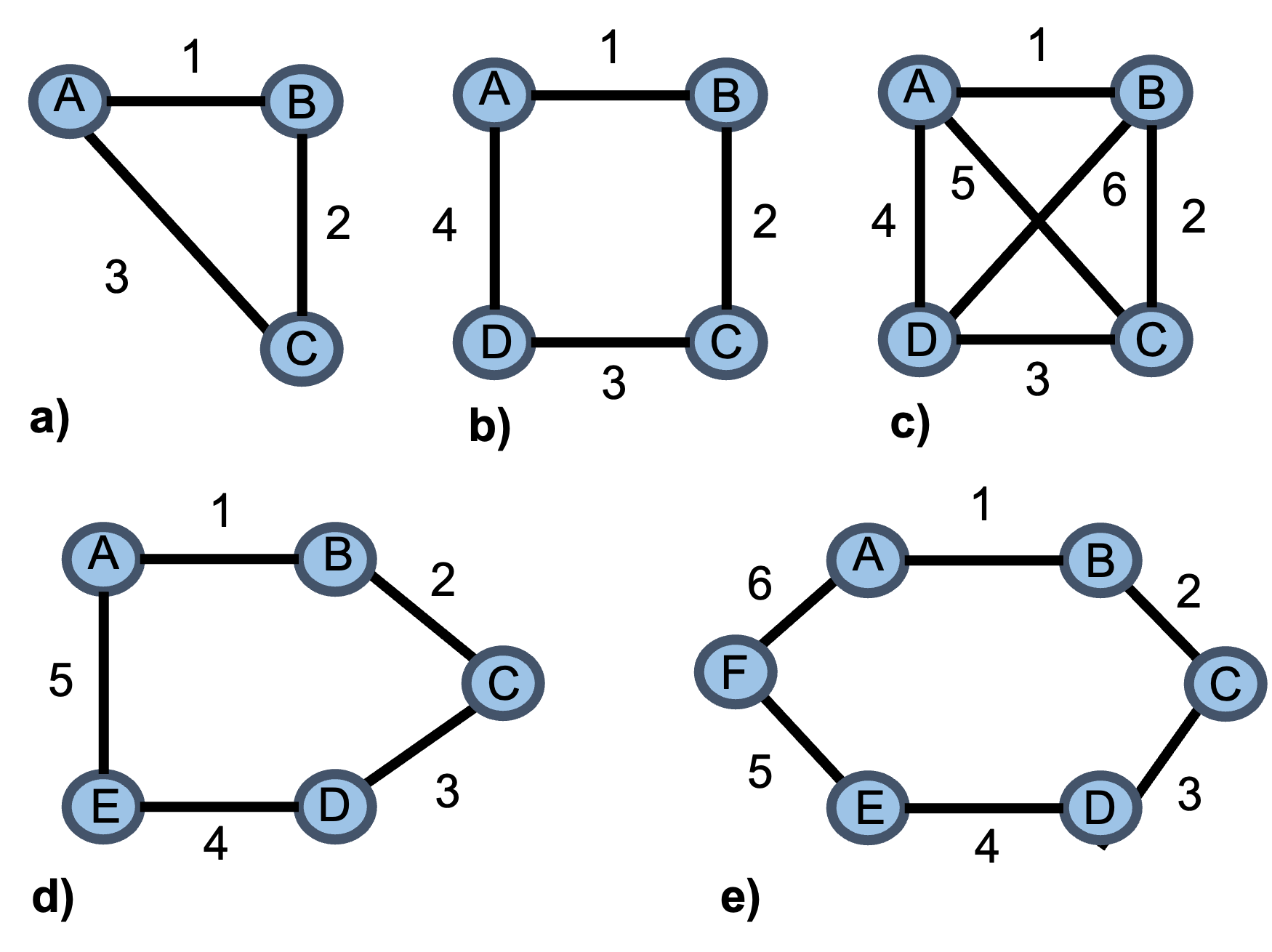}
    \caption{Various type of graphs used for simulation as benchmarks.
    }
    \label{5}
\vspace{-4mm}
\end{figure}

\begin{figure}
    \centering
    \includegraphics[width= 3.25in]{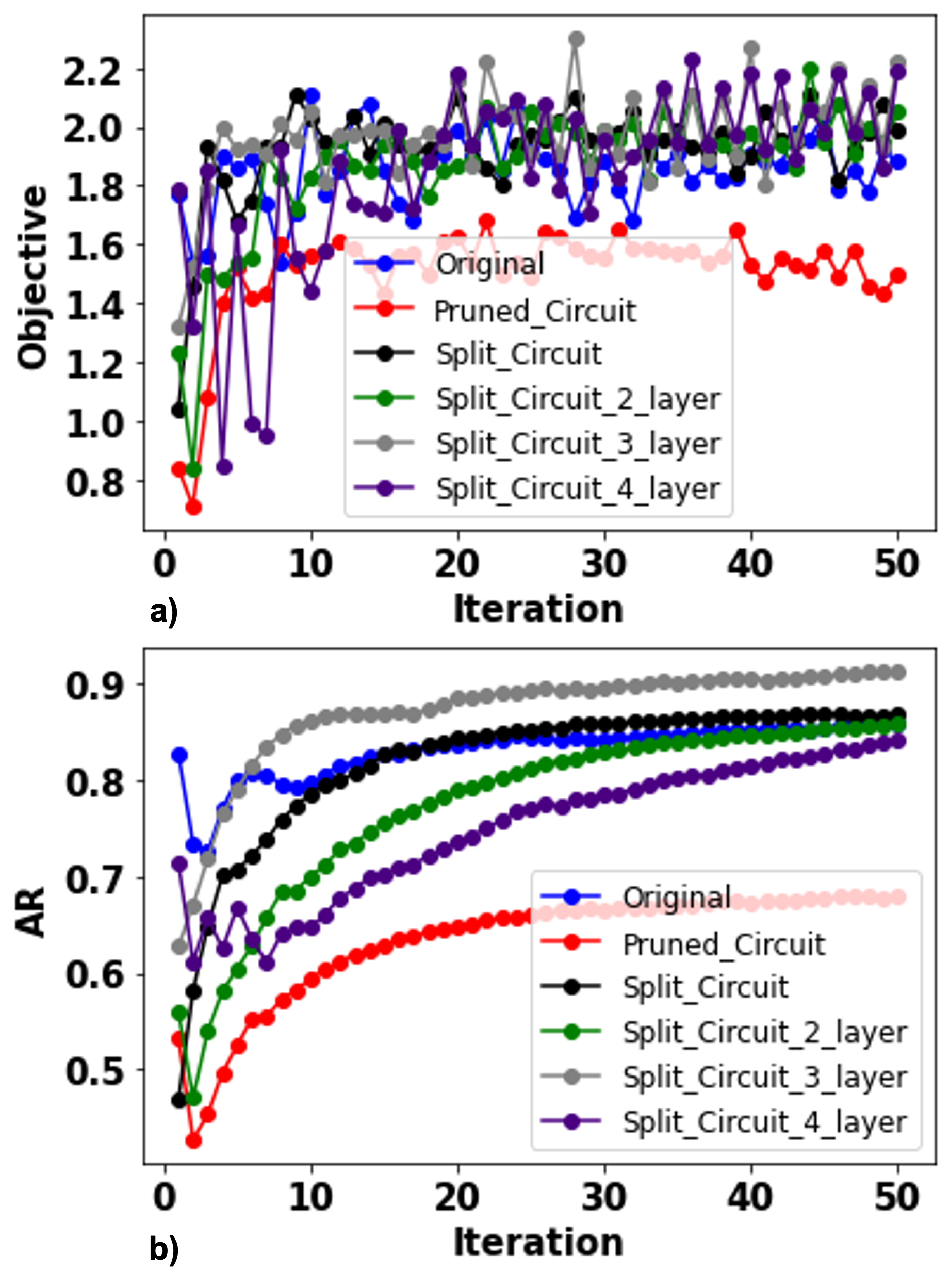}
    \caption{a) Objective and b) Approximation ratio (r) variation for a 3 node graph: (1) original circuit run on a single hardware (HW1), (2) circuit with one edge pruned run on a single hardware (HW1), (3) pruned circuits run on 2 different hardware, HW1 (25 iterations) and HW2 (25 iterations) in alternative fashion. }
    \label{8}
\vspace{-4mm}
\end{figure}

\begin{figure}
    \centering
    \includegraphics[width= 3.25in]{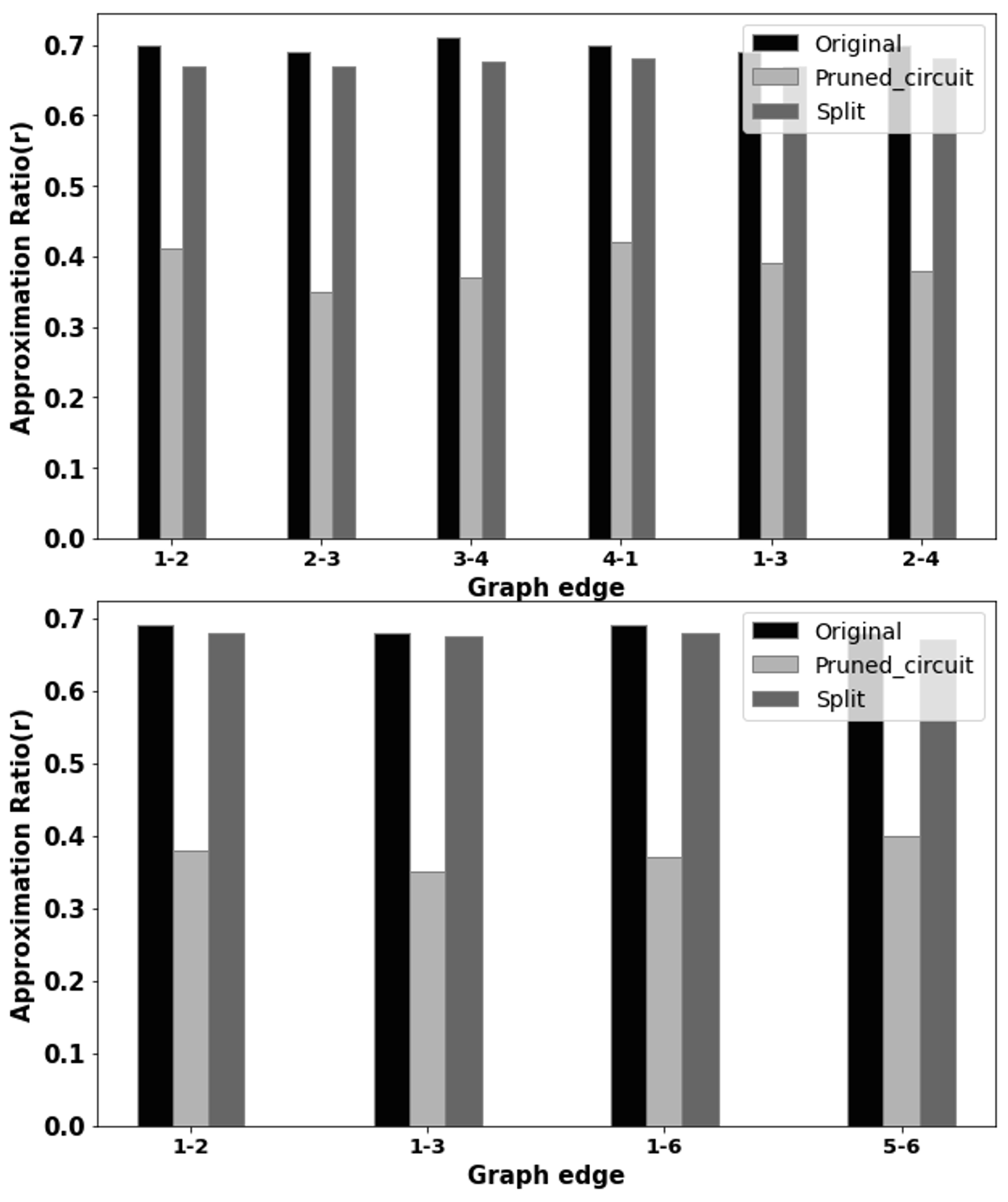}
    \caption{a) AR comparison for the 4 node graph (from Fig. \ref{5}b) between the original circuit run on a single hardware (HW1) vs a pruned circuit run on a single hardware (HW1) vs  the pruned circuits run on 2 different hardware, HW1 (25 iterations) and HW2 (25 iterations) in alternative fashion. b) Comparing AR using various combinations of single edge pruned for a fully connected 4 node graph (from Fig. \ref{5}c).
    }
    \label{6}
\vspace{-4mm}
\end{figure}

\subsection{Feasibility}

The proposed attack model is a feasible threat since, a) the cost of quantum computing is still high and many customers rely on cloud-based services to access these resources. We conducted research on the prices charged by various providers such as AWS Braket, IBM, and Google Cloud for access to their quantum processors, and found that prices range from  $\$$0.35 to  $\$$1.60 per second for qubit counts between 8 and 40. This cost can be significant for certain applications, such as factoring a 2048-bit product of two primes, which would require approximately 25 billion operations in 14238 logical qubits\cite{b37}. As more vendors enter the quantum service market, it is likely that some of these vendors will be untrusted, and may offer access to quantum hardware via the cloud at a lower cost, enticing users to use their services. This is more likely if the third-party vendor is based offshore, where labor, fabrication, packaging and maintenance costs are cheaper, b) accessing the quantum computers requires long wait time. When a user submits a job to a quantum system, it enters a scheduler where it is queued. The rapid expansion of quantum computing requirements has increased competition for these already scarce resources. For IBM Quantum systems \cite{b39}, reports indicate that only about 20$\%$ of total circuits have ideal queuing times of less than a minute, with an average wait time of about 60 minutes. Furthermore, more than 30$\%$ of the jobs have queuing times of more than 2 hours, and 10$\%$ of the jobs are queued for as long as a day or longer. This can be a significant barrier for certain applications, such as quantum machine learning, where quick access to the hardware is vital to lower the training and inference time. Third-party vendors may provide access to quantum hardware with little or no wait time, making it more desirable for users. However, this also poses a security risk as these vendors may not have the same level of security and controls in place as the established providers.

\section{Proposed Approach and Results}

In this section we present an overview of the proposed obfuscation procedure, followed by the results and analysis of the pruning-split obfuscation heuristic for solving the maxcut problem with QAOA for various graphs.

\subsection{Proposed Obfuscation Procedure}
We propose an obfuscation procedure for hiding the true functionality of a quantum circuit from untrusted hardware providers. Our approach involves pruning edges, or edge-specific gates, from the problem unitary of the circuit's problem graph. We demonstrate the effectiveness of this technique for QAOA to solve the maxcut problem. To protect the IP of the QAOA circuit, we eliminate one edge from the problem unitary for hardware 1 and include that edge but remove another for hardware 2. This creates two partial circuits that are run on different hardware with alternative iterations, making it challenging for either hardware provider to reverse engineer the full circuit. Fig. \ref{4} illustrates the proposed heuristic for a sample 4-node graph. To hardware 1, we send Circuit\_1 with the edge A-D removed and to hardware 2, we send Circuit\_2 with edge A-B removed making it challenging for either of the hardware provider to reverse engineer the full circuit. We use approximation ratio (AR) (Eq.\ref{1}) for benchmarking the performance of QAOA for the proposed obfuscation heuristic.

\begin{figure}
    \centering
    \includegraphics[width= 3.25in]{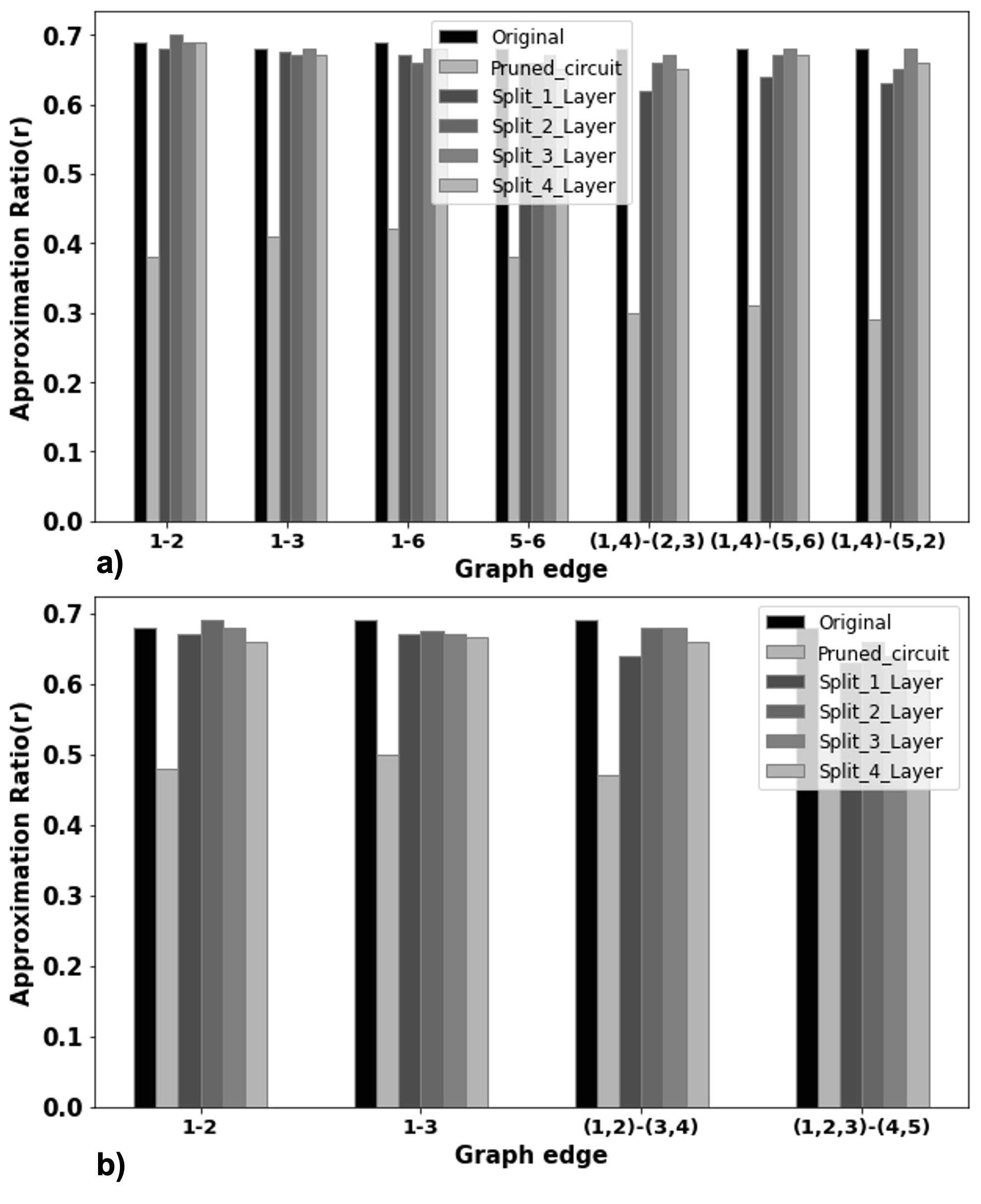}
    \caption{Performance variation with the number of layers used in the obfuscated and split cases. a) AR variation of the split circuit with different numbers of layers for a 4-node fully connected graph (from Fig. \ref{5}c) with various combinations of pruned edges. b) AR variation for a 5-node graph(from Fig. \ref{5}d) with various combinations of pruned edges.
    }
    \label{7}
\vspace{-4mm}
\end{figure}

\begin{table*}[]
    \centering
    \caption{Performance trends for various graphs (50 Iterations)}
    \begin{tabular}{ccc|ccccccccc}
    \hline
      &  & & & & Approximation ratio(r) & & & &\\ 
     \hline
      Graph & Pruned edge & Simulator & Original & Pruned circuit & Split & Split & Split & Split   \\
        & (HW1-HW2) & & & &(1-Layer) &(2-Layers)  & (3-Layers) &  (4-Layers)   \\
     \hline
    3(Fig.\ref{5}a) &  1-2 & Ideal & .81 & .62 & .78 & .89 & .89 & .9  \\
           &          & Fake\_backend & .79 & .59 & .77 & .84 & .84 & .83  \\
     \hline
     
    4(Fig.\ref{5}b) &  1-2 & Ideal & .74 & .53  & .71 & .84 & .86 & .9  &  \\
           &          & Fake\_backend & .7 & .42 & .67 & .73 & .75 & .72 \\
           & (1,4)-(2,3)  & Ideal & .73  &  .45 & .62  & .71  & .72 & .74  &  \\
           &          & Fake\_backend & .68 & .4  & .62 & .66 & .66 & .65 \\
        
     \hline
    4(Fig.\ref{5}c) &  1-2 & Ideal & .72  & .45  &.71  &.73  &.72 &.74  \\
           &          & Fake\_backend & .69 & .4  & .68  & .7 & .69 & .67 \\
           & (1,4)-(2,3)  & Ideal &.7  & .4 &.61  &.69  &.7 & .71   \\
           &          & Fake\_backend &.68  & .32 & .62  & .66  & .67 & .66  &  \\
           & (1,4)-(5,6)  & Ideal & .71 & .41 &.65  & .7  & .71 & .71  \\
           &          & Fake\_backend & .69  & .33  & .64 & .67 & .68 & .67 \\
     \hline
    5(Fig.\ref{5}d) &  1-2 & Ideal & .7 & .52  & .69  & .72 & .73 & .73  \\
           &          & Fake\_backend & .68  & .48  & .67 & .69 & .68 & .66 \\
           & (1,2)-(3,4)  & Ideal & .71 & .5 & .61 & .69 & .7 & .71 \\
           &          & Fake\_backend & .69  & .46 & .64  & .68  & .68 & .66  \\
           & (1,2,3)-(4,5)  & Ideal & .72  & .49 & .6 & .68 & .68 & .7  \\
           &          & Fake\_backend & .68 & .45  & .63  & .66 & .64 & .64 \\
    \hline
    6(Fig.\ref{5}e) &  1-2 & Ideal & .71 &.55 &.69 & .69  & .72 & .73   \\
           &          & Fake\_backend & .68 & .38 & .66 & .68 & .67 & .65  \\
           & (1,2)-(3,4)  & Ideal & .72  & .49 &.66 & .73 & .72 & .72  \\
           &          & Fake\_backend & .69 & .38 & .64 & .64 & .65 & .64  \\
           & (1,2,3)-(4,5)  & Ideal &.71  & .48 & .64 & .66 & .68 & .67 \\
           &          & Fake\_backend & .68  & .36 & .59 & .63 & .63 & .61 \\
           & (1,2,3)-(4,5,6)  & Ideal & .71 & .41  & .61 & .68 & .67 & .67 \\
           &          & Fake\_backend & .68 & .37 & .59 & .64 & .64 & .64 \\
           & (1,2,3,4)-(5,6)  & Ideal & .71   & .38 &.6  & .62 &.64 & .66 \\
           &          & Fake\_backend &.67  & .36 & .55 & .61 & .62 & .64  \\
     \hline
     \end{tabular}
    \label{tab:1}
\end{table*}

\subsection{Results and Analysis}

\subsubsection{Benchmark and Simulator}

We use the open-source quantum software development kit from IBM (Qiskit) \cite{b26} for simulations. We implement QAOA\cite{b14} the iterative algorithm to solve the combinatorial optimization problem MaxCut\cite{b24}. The MaxCut problem involves identification of a subset S$\in$V such that the number of edges between S and it's complementary subset is maximized for a given graph G = (V, E) with nodes V and edges E. Though MaxCut is an NP-hard problem\cite{b24}, there are efficient classical algorithms that can approximate the solution within a certain factor of optimality\cite{b25}. Using a p-level QAOA, an N-qubit quantum system is evolved with H\_C and H\_B p-times to find a MaxCut solution of an N-node graph Fig. \ref{4}. QAOA-MaxCut iteratively increases the probabilities of basis state measurements that represent larger cut-size for the problem graph. Qubits measured as 0’s and 1’s are in two different segments of the cut Fig. \ref{4}. Various sparse and dense graphs used for benchmarking are depicted in Fig. \ref{5}. We use the fake provider module in Qiskit as noisy simulators to run our benchmarks, which includes providers and backend classes that mimic real IBM Quantum systems using system snapshots. These snapshots contain important information about the quantum system, such as the coupling map, basis gates, and qubit parameters.

\subsubsection{Impact of edge selection on AR}

We used fake backends, HW1:fake\_montreal and HW2:fake\_mumbai, to run QAOA for the different graphs shown in Fig. \ref{5}, each for 50 iterations. We first ran the original representative circuit and compared it's performance to the proposed edge pruning and split technique. We exhaustively tried various combinations of pruning a single edge to obtain circuit\_1 run on HW1 and circuit\_2 run on HW2 for the graphs presented. Fig. \ref{8} showcases the objective and approximation ratio (r) variation for a 3 node graph. Here, the possible combinations are A-B: {1-2, 2-3, 3-1}, where A represents the edge removed for circuit\_1 and B represents the edge removed from the original circuit to obtain circuit\_2. Fig. \ref{6}a) shows the comparison for the 4 node graph (from fig. \ref{5}b)) between the original circuit run on a single hardware (HW1) vs the pruned circuits run on 2 different hardware, HW1 (25 iterations) and HW2 (25 iterations) in alternative fashion, keeping the overall iterations to 50. We observed that picking any combination had no significant effect on the final AR performance for the 4 node graph. We verified this result further by repeating the same for the 4 node fully connected graph from Fig. \ref{5}c). Using the results from the previous graph, we just compared one case from the distinct types of combinations possible such as a) one with an adjacent edge pruned [1-2,2-3,3-4,1-4], b) one with an opposite edge pruned [1-3,2-4], c) one regular edge and one edge from the diagonal [1-5,2-5,3-5,4-5,1-6,2-6,3-6,4-6] and d) both diameter edges {5-6}. We observe (Fig. \ref{6}b) that pruning any edge, irrespective of the connectivity, resulted in similar AR performance.

\subsubsection{Impact of pruning more than one edge at a time}

Pruning multiple edges simultaneously can hinder an adversary's ability to reverse engineer the circuit by limiting the amount of information available to them. However, AR degradation is much more significant. We observed to regain AR, additional layers can be added to the circuit, which will bring the AR performance closer to the original circuit (average recovery $\approx90\%$), but this also increases the execution time overhead. Table \ref{tab:1} shows the AR variation with a few different combinations of multiple edge pruned for various node graphs. For simplicity we only pruned adjacent edges and compared performance with varying number of layers.

\subsubsection{Performance comparison with varying QAOA layers}

Fig. \ref{7} represents the performance variation with the number of layers used in the obfuscated and split cases. We compare the performance with a single layer of the original circuit as the baseline. Fig. \ref{7}a) showcases the performance variation of the split circuit with different numbers of layers for a 4-node fully connected graph with various combinations of pruned edges, while Fig. \ref{7}b) depicts the same for a 5-node graph. We observe that for just one edge pruned in either case, the AR degradation when compared to the original is minimal, and increasing the number of layers may slightly improve AR but comes with a cost of computation time and resources overhead. For removal of 2 or more edges, the AR degrades significantly, which can be recovered by increasing the number of layers. However, increasing the number of layers past 2 leads to degraded performance, bounded by the qubit quality metrics (noise, short lifetime, and imperfect gate operations) of the target hardware. As per \cite{b18}, the optimal number of stages (p-value) for any QAOA instance is limited by the noise characteristics (gate error, coherence time, etc.) of the target hardware, as opposed to the current perception that higher-depth QAOA will provide monotonically better performance for a given problem compared to low-depth implementations.

\subsubsection{Performance comparison in ideal simulator vs fake back-ends}

The performance of QAOA on different node graphs is compared on ideal and fake backend simulators (Table-\ref{tab:1}). The results show that the average relative (AR) for each case is degraded by 10\% for the fake backend when compared to the ideal counterpart. As the number of layers in QAOA increases, there is an observed improvement in performance for all simulated graphs in the ideal simulator. However, when using the fake backend, after a certain point (depending on the number of nodes pruned), there is no increase or even some performance degradation with increasing the number of layers. This can be attributed to the hardware-induced errors that can alter the parameter and solution landscape of QAOA, limiting the performance gain achievable with higher numbers of layers. Furthermore, as the number of layers in QAOA increases, the circuit execution time also increases, which may exceed the qubit decoherence time and introduce more errors. In contrast, QAOA instances on ideal hardware have small variations between them.

\subsubsection{Adversarial reverse engineering effort}

The removal of a single edge from an n-node graph can significantly alter the properties of the original graph (and the problem encoded), making it difficult to determine the original structure. In the context of circuit obfuscation, an adversary's goal is to identify and add the removed edge(s) to retrieve the original graph/circuit. The number of trials required to determine the original graph depends on the properties of the original graph and the specific edge(s) that were removed. For example, consider an n-node complete graph, where every node is connected to every other node. In this case, any edge that is removed would be a bridge edge, and it would split the graph into two connected components, each with $(n-1)$ nodes. This means that the disconnected components would have a fixed number of edges, and it would be relatively easy to determine the original graph, as there is only one possibility. However, for an n-node cycle graph, where every node is connected to exactly two other nodes, forming a cycle if an edge is removed, the graph would become a tree, which is a connected graph with $(n-1)$ edges and $(n-1)$ nodes. In this case, the adversary would have $(n-1)(n-2)/2$ possible edges to check, and it would take $2^{(n-1)(n-2)/2}$ trials to determine the original graph in the worst-case scenario. Furthermore, the adversary has no way to validate the guess, and any attempt to reuse the circuit without adding the removed edge will result in corrupted or severely degraded performance. Adversary also lacks insights on the number of deleted edges which can add another layer of complexity. Therefore, the adversarial effort to reverse engineer the obfuscated design is high. For example, considering a 10-node graph. There can be $(10*9)/2 = 45$ possible edges that can be present or absent. For a complete graph with one deleted edge, the minimum number of guesses required is just 1. For a cyclic graph, the adversary would have (9*8)/2 = 36 possible edges to check which would take $2^{36} = 6.9 X 10^{10}$ trials to determine the original graph in worst-case scenario. 
In case the original graph is an arbitrary graph, the average number of guesses required would be the average of all possible graphs, which would be $2^{(45/2)} = 1.5X10^7$. The actual number of guesses required may vary depending on the specific graph and the specific edge that was removed.

\begin{figure}
    \centering
    \includegraphics[width= 3.25in]{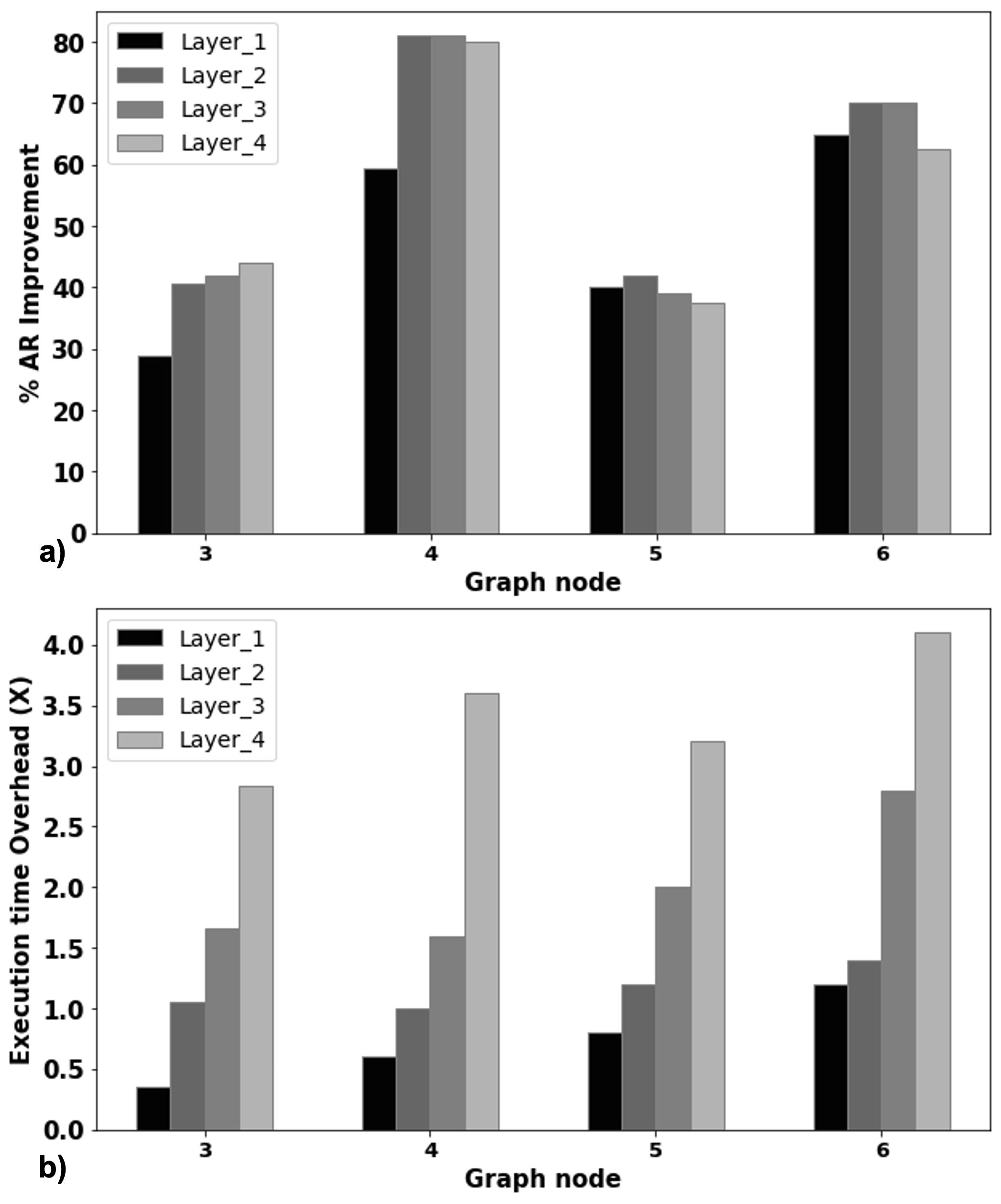}
    \caption{a) Improvement in AR with the number of layers, with one edge pruned for graphs (Fig. \ref{5}a,b,d,e). The baseline for comparison is a QAOA circuit with single pruned edge run on one untrusted hardware. b) Execution time for different layers compared to the baseline of a single edge pruned circuit.
    }
    \label{9} 
\vspace{-4mm}
\end{figure}

\subsection{Overhead analysis}

Using the proposed heuristic for the QAOA involves pruning edges which leads to a reduction in the performance (AR) of the algorithm. This reduction in performance can have a significant impact on the efficiency and effectiveness of the QAOA in solving optimization problems. However, for the cases when more than one edge is pruned performance can be nearly recovered to the original by increasing the number of layers at the cost of increased computation time and resources. Fig. \ref{9} compares the AR improvement versus execution time overhead with respect to the case where a single edge is pruned from a single layer QAOA circuit and is executed on a single untrusted hardware.  We observe an average improvement of 5\% in performance and AR recovery upto 90\% by increasing from layer 1 to layer 2 with an average time overhead increase of 0.4X. However, subsequent increases in the number of layers resulted in marginal performance improvement (0.1\% on average) at the cost of significant increase in time overhead (2x on average). This is a trade-off that needs to be carefully considered when implementing the proposed heuristic. Additionally, the use of the pruning and split heuristic requires an additional quantum hardware which can negate some of the cost savings from the use of cheaper quantum hardware. However, each hardware employs only half of the original number of iterations which can recover the cost due less computation time per hardware.

\subsection{Summary of findings}
\vspace{-1mm}

Following are the summary of results: (a) For pruning a single edge, any edge can be selected since it had little effect on the final AR performance using the proposed pruning and split heuristics. (b) Pruning multiple edges simultaneously degrades AR significantly. 
However, it can be recovered by adding additional layers of the circuit (average recovery $\approx 90\%$) at the cost of increased execution time by .4X on average. 
(c) Increasing the number of layers past 2 to recover AR after multiple edge removal may lead to poor performance (average AR degradation of $\approx8\%$ ) due to noise-induced qubit quality degradation 
of the target hardware. (d) If the original QAOA circuit contains multiple layers, the proposed addition of extra layers may lead to only minimal improvement (average 5\%). In such cases, the proposed pruning and split heuristic may lead to execution time overhead ($\approx 2X$ on average) to provide security guarantees.

\section{Discussion}

\subsection{Implementation on real hardware}

Fake backends can run on classical computers and are useful for testing and evaluating algorithms before they are run on actual hardware. Since fake backends are calibrated with real hardware on daily basis, the results obtained from fake backends are very realistic. Furthermore, simulation on fake backend is quick since it avoids the long wait queue associated with the real hardware. 
Running QAOA using the proposed pruning-split heuristic on actual hardware, such as a superconducting qubit or trapped ion quantum computer, can provide more accurate results however, the conclusions will remain the same. 

\subsection{QAOA vulnerabilities for other applications}

QAOA is primarily studied for use in quantum computing applications such as, quantum machine learning and quantum cryptography. However, as it is being proposed for use in fields like portfolio management, weather forecasting, and logistics, it is essential to consider potential vulnerabilities. In weather forecasting, QAOA can be used to optimize the forecast by finding the most likely state of the atmosphere given a set of observations. By learning about the number of observables and their inter-relationships from the QAOA circuit, an adversary can craft further attacks to manipulate the forecast, disrupting services for individuals, businesses, and governments that rely on accurate weather forecasts for decision-making. 
Similarly, in portfolio management, QAOA can be used to optimize investment decisions by finding the optimal combination of assets that maximize returns while minimizing risk. An adversary can learn about the assets from the QAOA circuit and design further attacks to either manipulate financial data or disrupt the optimization process, causing financial losses for investors using the portfolio management system. Additionally, an adversary may use this information to gain an unfair advantage in the financial markets by making profitable investments. 

\subsection{Consideration for adding an extra edge for obfuscation}
Adding an extra edge can also be one potential way for obfuscation in QAOA that could provide a level of protection for IP by making it more difficult for unauthorized parties to reverse engineer the algorithm and access sensitive information. However, this approach will increase the complexity and increased computational resources due to processing of an extra edge. Furthermore, the quality degradation due to cutting of extra edge/s could be difficult to recover in contrast to pruning of edge/s where recovery involves iteration with another circuit that has a different edge pruned. 

\subsection{Consideration for using more than 2 flavors of pruned circuit and multiple hardware}

Using multiple ``flavors" of pruned circuits, as well as multiple hardware providers, is one way to further mitigate potential risks.
In practice, this could be achieved by creating multiple versions of a circuit, each with a distinct set of edges eliminated, and alternating between these versions during the optimization process. Additional layers may be added to the circuit to recover optimization quality degradation caused by the pruning. It is important to note that this approach may result in performance degradation and added overhead costs, but it is a trade-off for increased security and privacy.

\section{Conclusion}

QAOA circuit can reveal the problem information to untrusted or unreliable hardware and pose a risk for IP theft. We propose an edge pruning obfuscation method for QAOA along with a split iteration methodology. This heuristic aims to secure IPs by running circuits containing partial information on different hardware, making it difficult for an adversary to access complete information. We also propose to increase the number of layers in QAOA circit to further mitigate the quality degradation associated with the proposed obfuscation. The effectiveness of this method is demonstrated on QAOA to solve graph maxcut problem of varying complexities. We show that the pruning-split methodology can impose factorial time reconstruction complexity on the adversary while incurring minimal overhead on solution quality.

\end{document}